\documentclass[aps,twocolumn,pra, 
superscriptaddress ]{revtex4-1}
\usepackage{braket,amsmath,amssymb,mathrsfs,dsfont}
\usepackage{tikz}
\usetikzlibrary{matrix,decorations.pathreplacing, calc,
positioning}
\usepackage{color}
\usepackage{graphicx}
\usepackage{dcolumn}
\usepackage{bm}

\usepackage{xcolor}
\definecolor{maroon}{RGB}{150,0,0}
\definecolor{dblue}{RGB}{0,0,150}
\usepackage[colorlinks=true,linkcolor=dblue,citecolor=maroon,urlcolor=maroon]{hyperref}
\begin{document}
\title{Experimental demonstration of 
the violation of the temporal Peres-Mermin inequality using
contextual temporal correlations and noninvasive measurements}
\author{Dileep Singh}
\email{dileepsingh@iisermohali.ac.in}
\affiliation{Department of Physical Sciences, Indian
Institute of Science Education \&
Research Mohali, Sector 81 SAS Nagar,
Manauli PO 140306 Punjab India.}
\author{Arvind}
\email{arvind@iisermohali.ac.in}
\affiliation{Department of Physical Sciences, Indian
Institute of Science Education \&
Research Mohali, Sector 81 SAS Nagar,
Manauli PO 140306 Punjab India.}
\author{Kavita Dorai}
\email{kavita@iisermohali.ac.in}
\affiliation{Department of Physical Sciences, Indian
Institute of Science Education \&
Research Mohali, Sector 81 SAS Nagar,
Manauli PO 140306 Punjab India.}
\begin{abstract}
We present a generalized quantum scattering circuit which can be used to
perform non-invasive quantum measurements, and implement it on NMR
qubits. Such a measurement is a key requirement for testing temporal
non-contextual inequalities. We use this circuit to
experimentally demonstrate the violation of the Peres-Mermin inequality
(which is the temporal analog of a Klyachko-Can-Binicioglu-Shumovsky
(KCBS) inequality), on a three-qubit NMR quantum information processor.
Further, we experimentally demonstrate the violation of a transformed
Bell-type inequality (the spatial equivalent of the temporal KCBS
inequality) and show  that its Tsirelson bound is the same as that for
the temporal KCBS inequality. In the temporal KCBS scenario, the
contextual bound is strictly lower than the quantum temporal and
nonlocal bounds.  
\end{abstract} 
\pacs{03.65.Ud, 03.65.Ta, 03.67.Ac, 03.67.Lx}
\maketitle
\section{Introduction} 
\label{intro} 
Intrinsic quantum correlations are used to distinguish between the quantum and
classical realms and are an important resource for quantum information
processing~\cite{nielsen-book-10}. The Bell inequality was proposed in 1964, to
provide bounds on classical correlations, and its violation implies
inconsistency of quantum mechanics with  locally realistic hidden variable
models~\cite{bell-ppf-64}.  In a different direction to identify intrinsic
quantumness, Kochen and Specker showed that quantum mechanics is contextual in
the sense that it does not come under the purview of noncontextual hidden
variable theories~\cite{ks-jmm-67}.  Quantum contextuality is a fundamental
quantum property of nature, which refers to the fact that the outcomes of
values of an observable can depend on the context provided by all other
compatible observables which are being measured along with
it~\cite{roy-pra-1993}.  Later work showed that quantum contextuality can  be
revealed by the violation of noncontextuality
inequalities~\cite{cabello-pra-13}.  It was shown that the Hardy-type and
GHZ-type proofs of the KS theorem involves a minimum of eighteen vectors for
any dimension, thereby verifying an old conjecture by
Peres~\cite{peres-prl-2020}.  Recently, a non-contextual hidden variable model
consistent with the kinematic predictions of quantum mechanics was
proposed~\cite{arora-pla-2019}, the set of quantum correlations that are
possible for every Bell and Kochen-Specker type contextuality was derived using
graph theory~\cite{correlation-pra-2019}, and the role of contextuality in
quantum key distribution (QKD) was explored~\cite{singh-pra-17}.

Klyachko-Can-Binicioglu-Shumovsky (KCBS) first proposed a
state-dependent inequality to test noncontextuality of
quantum correlations on a  single qutrit (three-level
indivisible quantum system)~\cite{KCBS}.  
Since then there have been several
state dependent and state independent proposals to test
contextuality~\cite{kurzynski-pra-12,cabello-prl-15,sohbi-pra-16,jaskaran-pra-17}.
Experimental tests of quantum contextuality have been
performed using
photons~\cite{nagali-prl-12}, trapped
ions~\cite{kirchmair-nature-09,leupold-prl-18},
and nuclear
spin qubits~\cite{dogra-pla-16,dileep-pra-19}.  The original
KS theorem was further extended to state independent
inequalities and three experimentally testable inequalities
were given which are valid for any noncontextual hidden
variable theory and can be violated by any quantum
state~\cite{cabello-prl-2008}.  
A
state-independent test of contextuality was designed by
Peres~\cite{peres-pla-1990} and by
Mermin~\cite{mermin-prl-1990}, which used a set of nine
dichotomic observables and involved
compatible measurements on them.  This inequality, called
the Peres-Mermin (PM) inequality, is considered the simplest
proof of the KS theorem for a four-dimensional Hilbert space
and relies on the construction of a Peres-Mermin square
with elements of the square being combinations of Pauli
measurements.

Bell-type inequalities are violated by quantum correlations
that exist between spatially separated sub-systems.  An
inequality to identify the intrinsic quantumness of temporal
correlations, known as the Leggett-Garg (LG) inequality,
assuming macroscopic realism and noninvasive measurements
was constructed~\cite{LG}.  Such temporal quantum
correlations can be revealed via noncommuting sequential
measurements on the same system at different times.  Later,
generalized multiple-measurement LG inequalities were
constructed and were interpreted using graph
theory~\cite{avis-pra-2010}.
Temporal quantum correlations have also been posited to be a
useful resource for quantum information processing protocols
and recently a theoretical framework for unifying spatial
and temporal correlations has been
developed~\cite{costa-pra-2018}.  Extensions of LG-type
nonlocal realistic inequalities have been studied in the
context of unsharp
measurements~\cite{dipankar-pra-2011,dipankar-pra-2015}. 
A recent scheme demonstrated that, temporal contextuality
which is generated from sequential projective measurements,
can be tested by violation of the KCBS
inequality~\cite{pan-ijtp-2019}.  
The structure of
temporal correlations for a single-qubit system was
characterized and experimental implementations on
nitrogen-vacancy centers in diamond were
explored~\cite{hoffmann-njp-2018}.  The genuine multipartite
nature of temporal correlations was confirmed by their
simultaneous violation of pairwise temporal
Clauser-Horne-Shimony-Holt (CHSH)
inequalities~\cite{ringbauer-npj-2018}.  The Tsirelson bound
refers to the maximum degree upto which a Bell inequality
can be violated~\cite{tsirelson} and is always less than the
algebraic bound~\cite{tsirelson-prl-2007,fritz-njp-2010}.
Surprisingly for LG-type inequalities, it was found that the
maximum degree to which the inequality can be violated is
greater than the Tsirelson bound, and the violation
increases with system size~\cite{budroni-prl-2014}.  It is
now well understood that the Bell theorem, the KS theorem
and the LG inequality are manifestations of the same
underlying hypothesis, namely, that quantum mechanics
contradicts noncontextual hidden variable (NCHV) theories of
physical reality.  A framework was developed to convert a
contextual scenario into equivalent temporal LG-type and
spatial Bell-type inequalities~\cite{markiewicz-pra-14}.

Temporal noncontextuality inequalities
typically require noninvasive
measurements to capture temporal quantum correlations, a
task not easy to perform experimentally.  State-independent
temporal noncontextuality inequalities were constructed and
used to obtain lower bounds on the quantum dimension
available to the measuring device~\cite{otfried-pra-14}. 
It was shown that
for measurements of dichotomous variables, the three-time LG
inequalities cannot be violated beyond the Luders bound,
which is numerically the same as the Tsirelson bound obeyed
by Bell-type inequalities~\cite{halliwell-pra-2020}.
Violations of LG
inequalities have been experimentally demonstrated using
polarized
photons~\cite{goggin-pnas-2011,dressel-prl-2011},
atomic ensembles~\cite{budroni-prl-2015}, a hybrid
optomechanical system~\cite{marchese-jpb-2020}, NMR
systems~\cite{souza-njp-11,athalye-prl-2011,katiyar-pra-2013,katiyar-njp-2017},
and superconducting
qubits~\cite{huffman-pra-2017}.
Recently, two- and three-time LG inequalities were
experimentally implemented on an NMR system, using
continuous in time velocity measurement and ideal negative
measurement protocols~\cite{laflamme-pra-2019}.  
Generalizations of
LG tests have been proposed for Bose-Einstein condensates
and atom interferometers~\cite{rosales-pra-2018}.

In this work, we experimentally demonstrate the violation of
a temporal contextuality PM inequality on an NMR quantum
information processor, using three spin qubits.  We
generalize the  quantum scattering circuit for two-point
correlation functions given in Ref.~\cite{souza-njp-11} to
measure $n$-point correlation functions, wherein an
observable is measured sequentially in time.  Performing $n$
successive measurements allowed us to achieve a non-invasive
measurement, without disturbing the subsequent evolution of
the system.  Unlike other measurement protocols, our circuit
is able to measure the desired temporal correlations in a
single experimental run and does not require additional CNOT
and anti-CNOT gates.  The violation of the temporal
noncontextual inequality demonstrates the contextual nature
of a particular quantum state during its time evolution.
State independent contextuality was tested via the violation
of the temporal analog of the KCBS inequality, the temporal
PM inequality, which was experimentally demonstrated by
sequentially measuring the three-point correlation function
and determining the expectation values of joint
probabilities.  We also demonstrated the violation of a
transformed Bell-type inequality, which is the spatial
analogue of the temporal KCBS inequality.  We have also
experimentally demonstrated that the Tsirelson bound of the
transformed Bell-type inequality is the same as that of the
temporal KCBS inequality.  For KCBS-type scenarios, the
quantum contextual bound is strictly lower than the temporal
and the nonlocal bound.  The measured experimental violation
of the inequalities match well with the theoretically
predicted bounds, within experimental errors.

This paper is organized as follows: The generalized quantum
scattering circuit and its deployment in generating
$n$-point time correlation functions is described in
Section~\ref{section2}.  
Section~\ref{section3a} contains details of the 
NMR system and experimental parameters used for the implementation
of the scattering circuit.
Section~\ref{section3b} describes
the experimental demonstration of the violation of the
temporal PM inequality, while Section~\ref{section3c} 
contains details of the implementation of the transformed Bell-type
inequality on three NMR qubits.  This section also contains
the experimental demonstration of the Tsirelson bound of the
Bell-type inequality, proving its equivalence to the bound
for the temporal inequality.  Section~\ref{section4} offers
a few concluding remarks about the scope and relevance of
our work.
\section{Generalized Quantum Scattering Circuit to 
Generate Temporal Correlations}
\label{section2}
Noninvasive measurements which
do not disturb the subsequent evolution of a system are in
general not possible in quantum mechanics, however, they can
be carried out in certain specific circumstances.  Several
noncontextual inequalities such as the LG inequality or the
temporal Bell-type inequalities require noninvasive
measurements, to capture temporal
quantum correlation.  Experiments to carry out such
noninvasive measurements are typically nontrivial to design
and implement.  
We describe here our generalized quantum scattering
circuit aimed at carrying out noninvasive measurements 
which we will use to investigate the violation of
temporal contextuality inequalities.

The standard quantum scattering circuit consists of a probe
qubit (ancillary) and the system qubit(s).  
The  generalized quantum scattering circuit which we
have designed to compute $n$ point correlations functions
involves
performing $n$ successive noninvasive measurements on an $N$
qubit quantum system, using only one ancilla qubit as the
probe qubit.  The circuit measures the $n$-point correlation
function $ \langle O(t_1)O(t_2)...O(t_n) \rangle $, wherein
an observable is measured sequentially at time instants
$t_1, t_2,... t_n$.
\begin{figure} 
\centering 
\includegraphics[scale=1.0]{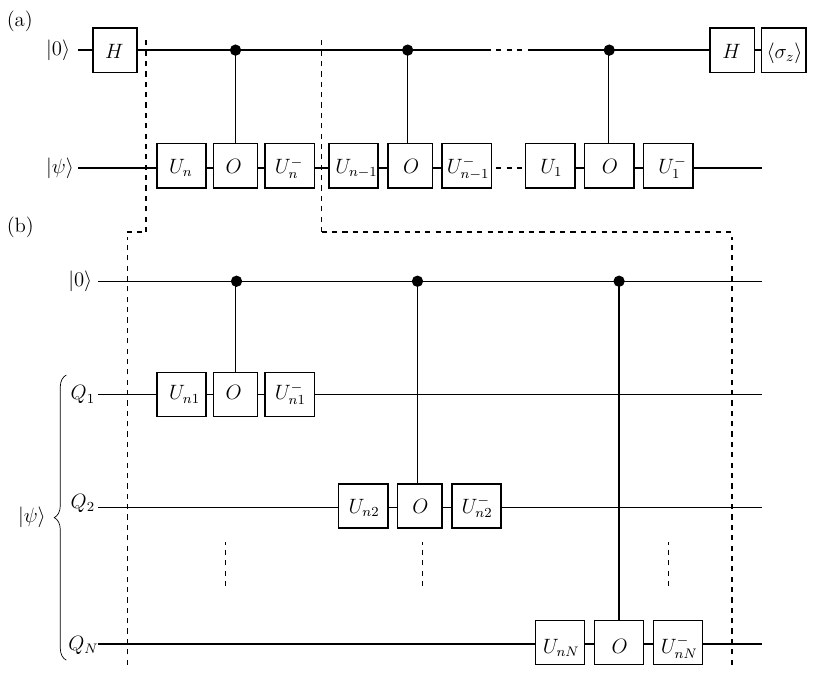}
\caption{ (a) Generalized quantum scattering circuit to measure the $n$
point time correlation function  $\langle [O_1(t_1) \otimes O_2(t_1)
\dots \otimes O_N(t_1)] [O_1(t_2) \otimes O_2(t_2) \dots \otimes
O_N(t_2)]\dots [O_1(t_n) \otimes O_2(t_n) \dots \otimes O_N(t_n)]
\rangle$, where each observable is a tensor product of $N$ operators,
$U^{\mp}_1=e^{\pm \frac{iHt_1}{\hslash }}$,...  $U^{\mp}_{n-1}=e^{\pm
\frac{iHt_{n-1}}{\hslash }}$, $U^{\mp}_{n}=e^{\pm
\frac{iHt_{n}}{\hslash }}$.The `probe' (ancilla) qubit is
initially in the state $\vert 0 \rangle$ and the system qubit is in the
state $\vert \psi \rangle$.  The correlation function is obtained by
measuring the the expectation value $\left \langle \sigma_z
\right\rangle$ of the ancilla qubit.  (b) Expanded schematic of the
circuit between dotted lines in panel (a), showing the decomposition of
the correlation function $\langle [O_1(t_n) \otimes O_2(t_n) \dots
\otimes O_N(t_n)] \rangle$, where $O_i(t_n)$ is measured on the $i$th
qubit ($i=1...N$) and $\vert \psi \rangle$ refers to the initial state
of all the system qubits and $U^{\mp}_{n1}=e^{\pm
\frac{iHt_{n1}}{\hslash }}$, $U^{\mp}_{n2}=e^{\pm \frac{iHt_{n2}}{\hslash
}}$..., $U^{\mp}_{nN}=e^{\pm \frac{iHt_{nN}}{\hslash }}$.}
\label{ckt-scatt-temp} 
\end{figure}

Fig.~\ref{ckt-scatt-temp} depicts a  schematic diagram 
of the generalized quantum
scattering circuit to generate temporal correlations and
demonstrate violation of temporal
noncontextuality.  
The system is prepared in a  known
initial state, which interacts with the
ancilla in such a way that a measurement
over its state after the interaction, brings out the information about the
system state.  
The `probe qubit' (ancillary qubit) is prepared in a known
initial state and the `system qubit' is prepared in the state for which
the observables are to be measured.  Consider the input state:
\begin{equation}
\rho_{in}= \rho_{probe}\otimes \rho_{sys}=\vert 0 \rangle 
\langle 0 \vert \otimes \vert \psi \rangle \langle \psi \vert
\end{equation}
where the `probe qubit' is prepared in the $\vert 0 \rangle $  state and the
`system qubit' is prepared in the state $ \vert \psi \rangle $.  After
applying the
unitary transformation shown in Fig.~\ref{ckt-scatt-temp}, 
the output is given by:
\begin{eqnarray}
&\rho_{{\rm out}}=\vert \psi_{{\rm out}} \rangle 
\langle \psi_{{\rm out}}\vert, \,\,{\rm with}  \nonumber \\
&\vert \psi_{{\rm out}} \rangle 
= \vert 0 \rangle \otimes (I+U)\vert \psi \rangle +
\vert 1 \rangle \otimes (I-U)\vert \psi \rangle,\,\, {\rm
and} \nonumber
\\
&U=e^{\frac{iHt_1}{\hslash }} O e^{-\frac{iHt_1}{\hslash }}
e^{\frac{iHt_{2}}{\hslash }} O e^{-\frac{iHt_{2}}{\hslash
}}\cdots
e^{\frac{iHt_n}{\hslash }} O e^{-\frac{iHt_n}{\hslash }} 
\end{eqnarray}
The real part of the
expectation value of the $z$-component of the spin
angular momentum of the `probe' qubit turns
out to be related to the expectation values of desired
observables of the
original state as follows: 
\begin{eqnarray}
& \langle
\sigma_z \rangle = Tr[\rho_{sys} U] \quad {\rm
therefore},\nonumber \\
&\langle \sigma_z \rangle = \langle O(t_1)O(t_2)...O(t_n) \rangle
\end{eqnarray}
The generalized quantum scattering circuit can be used to
experimentally demonstrate
those inequalities which involve temporal correlation functions, such as
the temporal PM noncontextual inequality and the temporal KCBS inequality.
While the ideal negative measurement (INM) protocol described in 
Ref.~\cite{laflamme-pra-2019} is similar to our measurement scheme, 
in the INM protocol the ancilla is coupled to
only one of the two measurement outcomes and
the protocol hence requires two experimental runs: with a CNOT gate as well
as with an anti-CNOT gate. Our circuit on the other hand, requires only
a single experimental run and does not require additional CNOT and anti-CNOT
gates for its implementation.
\section{Violation of Temporal PM and Temporal Bell-Type Inequalities}
\label{section3}
Consider performing a set of five dichotomic (i.e. the
measurement outcomes are $\pm 1$) measurements of 
variables
$X_j, j=1,..5$ on a single system.  Each measurement $X_j$
is compatible with the preceding and succeeding measurements
and the sums are modulo 5.  Compatible measurements implies
that the joint or sequential measurements of the variables
$X_j$ do not affect each other, which basically ensures that
the measurements are noninvasive.  The existence of a joint
probability distribution for all the measurement outcomes
can be tested by constructing the KCBS
inequality~\cite{markiewicz-pra-14}:
\begin{equation}
\sum_{j=0}^{4} \langle X_j 
X_{j+1} \rangle \ge - 3
\end{equation}
where $-3$ is the minimum value for an NCHV model.
Noncontextual in this sense implies that the
NCHV theory assigns a value to an observable
which is independent of other
compatible observables being measured
along with it.
By definition each correlation function is
given by~\cite{markiewicz-pra-14}:
\begin{equation}
\langle X_i X_j \rangle = \sum_{x_i,x_j=\pm 1}
x_i x_j p(x_i,x_j)
\end{equation}
\begin{figure}
\centering
\includegraphics[scale=1.0]{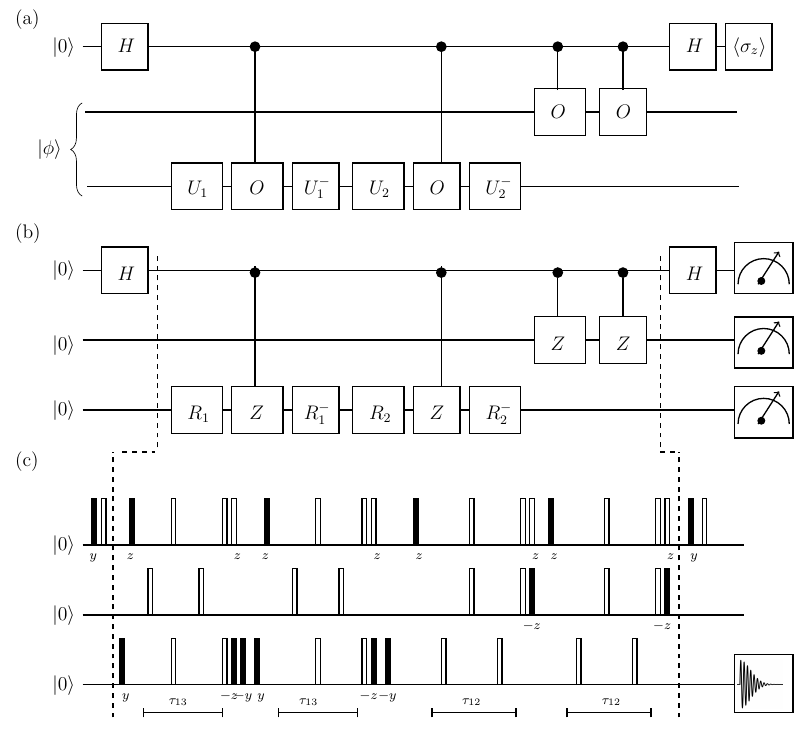}
\caption{(a) Quantum scattering circuit for measuring the correlation function
$\langle A \alpha a  \rangle$ involved
in the PM inequality, where $O = \sigma_z $ and $U^\pm_{1,2} =
e^{\mp i \sigma_y \theta/2}$ with $\theta = \pi/2$.
(b) Decomposition of the quantum scattering
circuit in terms of rotation operators where
$R^{\pm}_{1,2}$ correspond to 
$(\frac{\pi}{2})_{\pm y}$, $H$ are Hadamard 
gates and $Z$ are rotations about the $z$ axis.
(c) NMR pulse sequence corresponding to the quantum
scattering circuit, where filled and unfilled
rectangles correspond to $\pi/2$ and $\pi$ pulses,
respectively. The time intervals 
$\tau_{12},\tau_{13}$ are set to
$\frac{1}{2 J_{HF}}$ and
$\frac{1}{2 J_{HC}}$, respectively.
}
\label{scat-pm-8dim} 
\end{figure}

\begin{figure}
\centering
\includegraphics[scale=1.0]{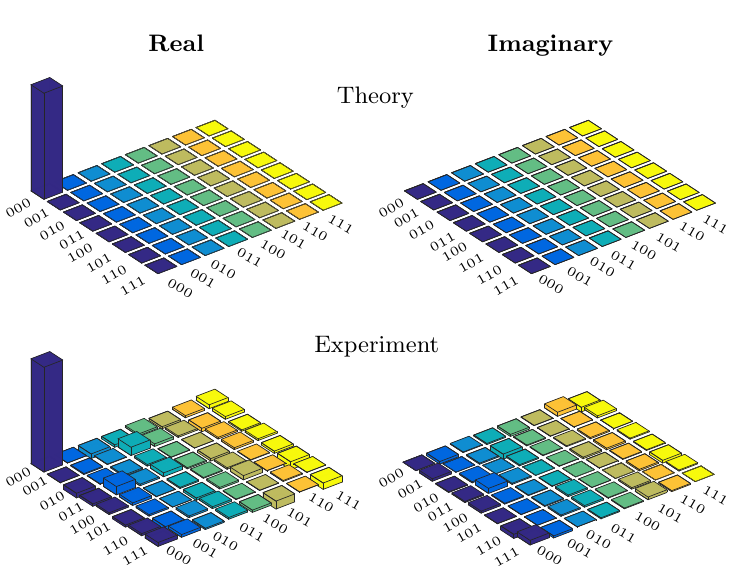}
\caption{Real (left) and imaginary (right) parts of the theoretical and
experimental tomographs of the  input $ \rho =\vert 0 \rangle \langle 0
\vert \otimes \vert 00 \rangle \langle 00 \vert  $  state in the
eight-dimensional Hilbert space, prepared with an experimental state
fidelity of 0.964$\pm$0.004.}
\label{tomo8dim}
\end{figure}

A ``pentagon LG'' inequality was constructed wherein~\cite{avis-pra-2010}
\begin{equation}
\sum_{1 \le i < j \le 5} \langle
X_i X_j \rangle + 2 \ge 0
\label{pentagon}
\end{equation}
This inequality has 10 two-time correlation functions
which can be computed from one single experiment using
compatible measurements.
The two-time correlation function turns out to 
be~\cite{markiewicz-pra-14}
\begin{equation}
\langle X_i X_j \rangle = \frac{1}{2} Tr[\rho \{X_i, X_j\}]
\end{equation}
for a density matrix $\rho$.
The five measurable observables were chosen to be~\cite{otfried-pra-14}:
\begin{equation}
X_1 \equiv \sigma_z\,,
X_2 \equiv \sigma_{\theta} \,,
X_3 \equiv \sigma_z \,,
X_4 \equiv \sigma_{\theta} \,,
X_5 \equiv \sigma_z 
\end{equation}
where $\sigma_x, \sigma_z$ are the Pauli operators
and $\sigma_{\theta} \equiv
                \cos{\theta}\, \sigma_z +
                \sin{\theta}\, \sigma_x$.
For this set of chosen observables
and with
$\theta$ chosen such that $\cos{\theta}=-3/4$,
the correlation function takes the value~\cite{avis-pra-2010}
\begin{equation}
\sum_{1 \le i < j \le 5} \langle X_i X_j \rangle = -9/4
\end{equation}
which is the smallest possible value
and  violates 
the ``pentagon'' LG inequality given in
Eqn.~(\ref{pentagon}).
\subsection{The NMR system}
\label{section3a}
We used the molecule of ${}^{13}$C -labeled diethyl
fluoromalonate dissolved in acetone-D6 as a three-qubit
system, with the ${}^{1}$H, ${}^{19}$F and ${}^{13}$C
spin-1/2 nuclei being encoded as `qubit one', `qubit two'
and `qubit three', respectively.  The NMR Hamiltonian for a
three-qubit system  in the rotating frame
is~\cite{oliveira-book-07}: 
\begin{equation}
\mathcal{H}=-\sum_{i=1}^{3} v_i I^i_z + \sum_{i>j,i=1}^{3} J_{ij} I^i_z I^j_z 
\end{equation}
where the indices $i, j$ = 1, 2, or 3 label the qubit,
$\nu_{i}$ is the chemical shift of the $i$th qubit in the
rotating frame, $J_{{ij}}$ is the scalar coupling
interaction strength, and $I_z^{{i}}$ is $z$-component of
the spin angular momentum operator of the $i^{th}$ qubit.
The system was initialized in a pseudopure state (PPS),
i.e., $ \vert 000 \rangle$, using the spatial averaging
technique~\cite{cory-spatial}.  The fidelity of the
experimentally prepared PPS state was computed to be
0.964$\pm$0.004 using the Uhlmann-Jozsa fidelity
measure~\cite{jozsa,uhlmann}.  Quantum state tomography was
performed to experimentally reconstruct the density operator
using a reduced tomography protocol~\cite{leskowitz}.  The
$T_1$ and $T_2$ relaxation times for all three qubits range
between 3.7 s - 6.8 s and 1.0 s - 2.8 s, respectively.
Nonlocal unitary operations were achieved by free evolution
under the system Hamiltonian, of suitable duration under the
desired scalar coupling with the help of embedded $\pi$
refocusing pulses.  The durations of the $\frac{\pi}{2}$
pulses for ${}^{1}$H, ${}^{19}$F, and ${}^{13}$C nuclei were
9.55 $\mu$s at 18.14 W power level, 23.00 $\mu$s at a power
level of 42.27 W, and 15.75 $\mu$s at a power level of
179.47 W, respectively.

\subsection{Experimental violation of the temporal Peres-Mermin inequality}
\label{section3b}
A temporal equivalent of the KCBS inequality can be constructed
similarly to the ``pentagon LG'' inequality by considering a set of nine
dichotomic variables, and three successive
measurements 
at two sequential times from the set of time points
$t = \{ t_1,t_2,..t_5 \}$.
The observable set chosen is the
``PM square'' of nine dichotomous and mutually
compatible observables 
$A,B,C,a,b,c,\alpha,\beta,\gamma$~\cite{otfried-pra-14}:
\begin{equation}
\begin{array}{lll}
A=\sigma_z \otimes I, & B= I \otimes \sigma_z,  &C=\sigma_z \otimes \sigma_z \\
a=I \otimes \sigma_x, & b=\sigma_x \otimes I, &  c=\sigma_x \otimes \sigma_x\\
\alpha =\sigma_z \otimes \sigma_x, &  \beta =\sigma_x
\otimes \sigma_z, &  \gamma =\sigma_y \otimes \sigma_y.
\end{array}
\end{equation}
Consider the combination of expectation values defined as
follows:
\begin{equation}
\langle X_{{\rm PM}} \rangle =
\langle ABC \rangle+
\langle bca \rangle+
\langle \gamma \alpha \beta \rangle+
\langle A \alpha a \rangle+
\langle b B \beta \rangle-
\langle \gamma c C \rangle
\end{equation}
If we make non-contextual assignments of values we get the
inequality
\begin{equation}
\langle X_{{\rm PM}} \rangle \le 4
\end{equation}
which is satisfied by all NCHV theories.
This is the temporal 
PM inequality ($X_{PM}$)~\cite{otfried-pra-14}.
It has been shown that for 
a four-dimensional
quantum system  and a particular set of observables,
a value of
$\langle X_{PM} \rangle = 6$ is obtained for
any quantum state, demonstrating state-independent
contextuality~\cite{cabello-prl-2008}.

We note here in passing that in this ``PM square'' set of
measurements, each observable always occurs either in the
first place or the second place or the third place in the
sequential mean value. This inequality is violated whenever
a joint probability distribution cannot be found which
assigns predetermined outcomes to the measurements $X_i$ at
all times $t_1...t_5$, and this violation is termed contextual
in time.  The system evolves under the action of a
time-independent Hamiltonian $ H = \hbar \omega
\sigma_{x,y}$, which can be implemented in NMR using
suitable rf pulses applied on the qubits.  After state
preparation, the probe qubit interacts with the system qubit
via suitable unitaries.  The temporal correlation functions
are obtained by measuring the real part of the expectation
value of $z$-component of the spin angular momentum of the
probe qubit.  

Our experimental task is to measure the expectation values
of joint probabilities which are measured sequentially.  To
violate the temporal PM inequality we need to measure the
three observables sequentially for any two-qubit state.  We
experimentally violated the PM inequality by measuring the
six correlation functions using the generalized quantum
scattering circuit.  Fig.~\ref{scat-pm-8dim} shows the
quantum scattering circuit, the operator decomposition
and the corresponding NMR pulse sequence, to calculate the correlation
function $\left \langle A \alpha a \right\rangle$ which is
one of the six correlation function used in the PM temporal
inequality. The PM temporal inequality is violated for any
two-qubit state. The probe qubit is prepared in
known $ \vert 0 \rangle  $ state and system qubit is
prepared in $ \vert \phi \rangle = \vert 0 0 \rangle    $
state. We apply the transformation given in
Fig.~\ref{scat-pm-8dim}(a), with suitable values of $ O =
\sigma_z $ and $ \theta = \pi/2 $. The correlation function
$\left \langle A \alpha a \right\rangle$ for the $ \vert
\phi \rangle = \vert 0 0 \rangle$ state can be obtained
by measuring the real part of the expected value of the
$z$-component of the spin for the probe qubit. The other correlation
functions involved in the PM temporal inequality are
measured in a similar fashion.

Since the temporal PM inequality is violated for any
two-qubit state, we chose to prepare the probe qubit in a known $
\vert 0 \rangle  $ state and the system qubits were prepared
in the $ \vert \phi \rangle = \vert 0 0\rangle$ state. The
experimental tomograph of the state prepared in  $ \rho
=\vert 0 \rangle \langle 0 \vert \otimes \vert 00 \rangle
\langle 00 \vert  $ is given in Fig~\ref{tomo8dim}, achieved
with a fidelity of 0.964$\pm$0.004.  We applied the unitary
transformations given in Fig.~\ref{scat-pm-8dim} with values
of $O=\sigma_z $ and $ \theta = \pi/2 $. The correlation
function $\left \langle A \alpha a  \right\rangle$ for the $
\vert \phi \rangle = \vert 0 0 \rangle    $ state can be
obtained by measuring the real part of the expected value of
the spin $z$-component of the probe qubit.  The other
correlation functions involved in the temporal PM inequality
are calculated in a similar fashion.  The mean value of the
correlation functions and their error bars were calculated
by repeating the experiment three times and the
theoretically expected and experimentally calculated values
are given in Table~\ref{3qubit-pm-table}.  The theoretically
computed and experimentally measured values of the
correlation functions agree well to within experimental
errors.  We experimentally violated the temporal PM
inequality, obtaining $\langle X_{PM}
\rangle_{\rm Expt}= 4.667\pm0.013$,  showing the contextual
nature of the measured expectation values.
\subsection{Experimental violation of a temporal Bell-type inequality}
\label{section3c}
\begin{table}
\caption{Theoretically computed and experimentally measured
values of  correlations functions corresponding to the PM
inequality .}  
\centering
\begin{tabular}{|c|c|c|}
\hline
Observables~ & ~Theoretical~  & 
~Experimental~\\
\hline
\hline
$ \langle ABC \rangle$ & 1  & 0.928 $\pm$ 0.017\\ 
\hline
 $ \langle bca \rangle$ &  1 & 0.706 $\pm$ 0.012 \\
\hline
$ \langle \gamma \alpha \beta \rangle$ & 1 & 0.817 $\pm$ 0.010\\
\hline
 $ \langle A \alpha a \rangle$ &  1  & 0.685 $\pm$ 0.008\\
\hline
 $ \langle b B \beta \rangle$ &  1  & 0.755 $\pm$ 0.011\\
\hline
 $ \langle \gamma c C \rangle$ &  -1 & -0.784 $\pm$ 0.019 \\
\hline
\hline
\end{tabular}
\label{3qubit-pm-table}
\end{table} 
The temporal KCBS noncontextual inequality 
can be constructed by considering a dichotomic variable $X_t$ with
successive measurements performed at two sequential times drawn
from the time instants $t = \{t_0,t_1,....,t_4\}$. The two-point
temporal correlations thus obtained lead to the corresponding temporal
KCBS inequality~\cite{markiewicz-pra-14}:
\begin{equation}
\sum_{i=0}^{4} \langle X_{t_i} X_{t_{i+1}} \rangle \ge -3
\end{equation}
The violation of this inequality can be termed as contextuality in time.

The temporal KCBS inequality
can be
transformed into a Bell-type inequality which tests the
existence of a joint probability distribution for 
measurements on dichotomic variables,
performed on
subsystems $A$ and $B$. The transformed Bell-type
inequality is given as~\cite{markiewicz-pra-14}
\begin{equation}
 \left \langle A_0 B_1  \right\rangle + \left \langle A_1
B_2  \right\rangle +  \left \langle A_2 B_3  \right\rangle +
\left \langle A_3 B_4  \right\rangle + \left \langle A_4 B_0
\right\rangle \geq -3 
\label{transformed-bell}
\end{equation}
where $A_i$ and $B_j$ are measured on the subsystems with the
additional constraint that 
\begin{equation}
\langle A_i B_i \rangle = 1  \,\,\, {\rm for \,\,\, all} \,\,\, i
\end{equation}
which implies that the outcomes of pairs of measurements are
the same.
Violation of this inequality shows the non-existence of joint probability
distribution for this scenario. 
\begin{figure}
\centering
\includegraphics[scale=1.0]{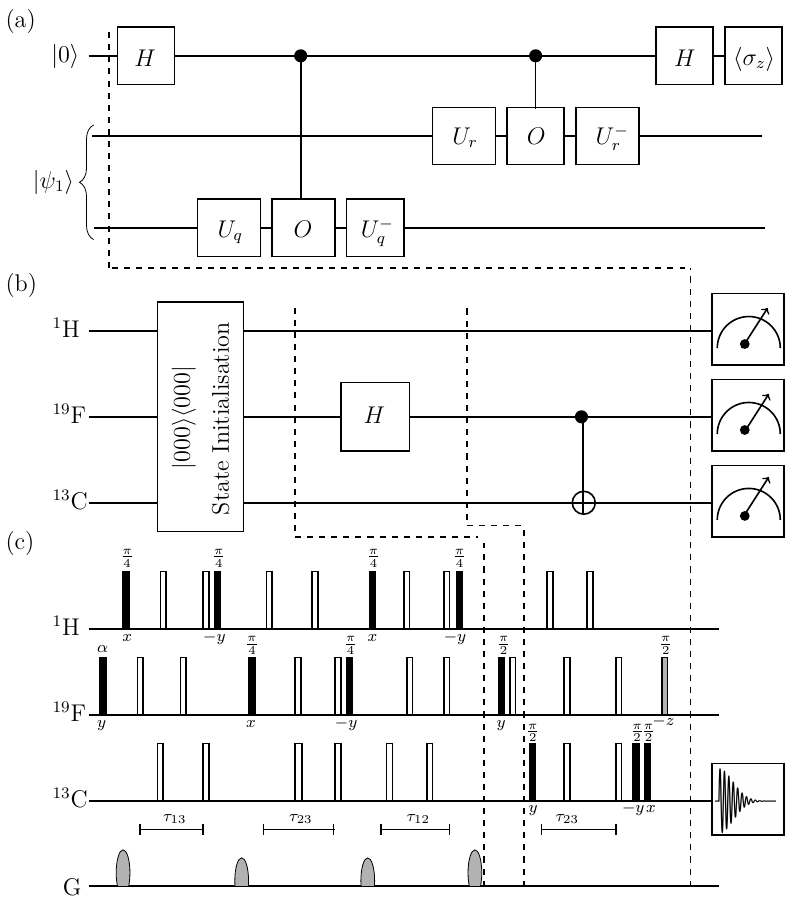}
\caption{(a) Quantum circuit to measure the correlation
function $\langle A_r B_q \rangle$ involved in the Bell-type
inequality, where $U_{r,q} = e^{\frac{- i 2 \pi r,q}{5}}$,
$O = \sigma_z$ and $r,q=0,1,2,3,4$.  (b) Quantum circuit for
state preparation. (c) Corresponding NMR pulse sequence for
the quantum circuit. The sequence of pulses before the first
dashed black line achieves initialization of the state into
the pseudopure $\vert 000 \rangle$ state. The unfilled
rectangles denote $\pi$ pulses, and the flip angle and
phases of the other pulses written below each pulse. The
time intervals $\tau_{12}$, $\tau_{13}$, $\tau_{23}$ are set
to $\frac{1}{2J_{HF}}$, $\frac{1}{2J_{HC}}$,
$\frac{1}{2J_{FC}}$, respectively.}
\label{ckt-pulse-bell}
\end{figure}
\begin{figure}
\centering
\includegraphics[scale=1.0]{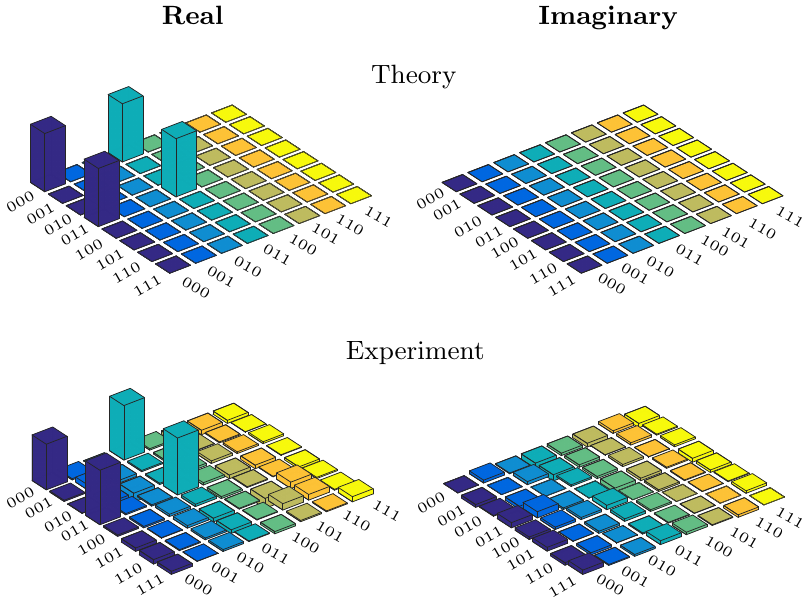}
\caption{Real (left) and imaginary (right) parts of the
theoretically expected and the experimentally reconstructed
tomographs of the  $\langle \psi_1 \vert =
\frac{1}{\sqrt{2}}(1,0,0,1,0,0,0,0) $ state in the
eight-dimensional quantum system, with an  experimental
state fidelity of 0.947$\pm$0.009.}
\label{tomo-bell-3q}
\end{figure}

We experimentally demonstrated the violation of the
transformed Bell-type inequality given in
Eqn.~\ref{transformed-bell} using the quantum scattering
circuit on the same three-qubit system.
Fig.~\ref{ckt-pulse-bell}(a) shows the quantum scattering
circuit to calculate the correlation function $ \langle A_r
B_q \rangle$, involved in the transformed Bell-type
inequality on an eight-dimensional quantum system.  For the
violation of the transformed Bell-type inequality, we used
the ${}^{1}$H as the probe qubit and ${}^{13}$C and
${}^{19}$F as the system qubits.  We apply the
transformations given in Fig.~\ref{ckt-pulse-bell}(a) with
suitable values of $ O = \sigma_z $ and $ q,r = 0,1,2,3,4 $.

The optimal violation of transformed Bell-type inequality can be
obtained for the state $\langle \psi_1 \vert =
\frac{1}{\sqrt{2}}(1,0,0,1)$ with the probe qubit prepared
in the state $\vert 0 \rangle$,
and for the
measurements $A_j= \sigma_j \otimes I $, $B_j= I \otimes
\sigma_j $ where $ j= 0, 1, 2, 3, 4 $ and  $\sigma_j=
e^{i\frac{2 \pi j}{5}\sigma_y} \sigma_z  e^{-i\frac{2 \pi
j}{5}\sigma_y}$.  The correlation functions $\langle A_r B_q
\rangle$  for the state $\langle \psi_1 \vert =
\frac{1}{\sqrt{2}}(1,0,0,1) $ can be obtained by
measuring the real part of the expected value of the spin
$z$-component for the probe qubit.  The corresponding
quantum circuit for state preparation is shown in
Fig.~\ref{ckt-pulse-bell}(b) and the NMR pulse sequence is
shown in Fig.~\ref{ckt-pulse-bell} (c). The sequence of
pulses before the first dashed black line achieves state
initialization into the $\vert 000 \rangle$ state. After
this we apply the Hadamard gate (on ${}^{13}$C), followed by
a CNOT$_{23}$ gate, and the resultant state corresponds to $
\rho_1 =\vert 0 \rangle \langle 0 \vert \otimes \vert \psi_1
\rangle \langle \psi_1 \vert  $ with $\langle \psi_1 \vert =
\frac{1}{\sqrt{2}}(1,0,0,1) $.

The tomograph of the state prepared in  $ \rho_1 =\vert 0
\rangle \langle 0 \vert \otimes \vert \psi_1 \rangle \langle
\psi_1 \vert  $  with $\langle \psi_1 \vert =
\frac{1}{\sqrt{2}}(1,0,0,1) $ is given in
Fig~\ref{tomo-bell-3q} with an experimental fidelity of
0.947$\pm$0.009.  The mean values of the correlation
functions and their error bars were calculated by repeating
the experiment three times and and calculated values are
given in Table ~\ref{3qubit-bell-table}. As seen from the
values tabulated in Table~\ref{3qubit-bell-table}, the
theoretically computed and experimentally measured values of
the correlation functions agree well to within experimental
errors.  We have experimentally violated the transformed
Bell-type inequality  with the
violation of $-3.755\pm0.008$.  
When a temporal and a spatial scenario are interconvertible,
the corresponding temporal and spatial Tsirelson bounds are
always equal and are greater than or equal to the contextual
Tsirelson bound~\cite{tsirelson}.
We also experimentally
verified  that the Tsirelson bound of the transformed
Bell-type inequality is the same as that of the temporal
KCBS inequality.  For KCBS-type scenarios, the quantum
contextual bound is strictly lower than the quantum temporal
and nonlocal bound.

\begin{table}
\caption{Theoretically computed and experimentally measured
values of quantum correlations corresponding to the
Bell-test.}  
\vspace*{12pt}
\centering
\begin{tabular}{|c|c|c|}
\hline
Observables~ & ~Theoretical~  & 
~Experimental~\\
\hline
\hline
$ \langle A_{0} B_{1} \rangle$ & -0.809  & -0.684 $\pm$ 0.014\\
\hline
 $ \langle A_{1} B_{2} \rangle$ &  -0.809  & -0.754 $\pm$ 0.006 \\
\hline
$ \langle A_{2} B_{3} \rangle$ &  -0.809  & -0.756 $\pm$ 0.011\\
\hline
 $ \langle A_{3} B_{4} \rangle$ &  -0.809  & -0.746 $\pm$ 0.005\\
\hline
 $ \langle A_{4} B_{0} \rangle$ &  -0.809 & -0.815 $\pm$ 0.004
\\
\hline
\hline
\end{tabular}
\label{3qubit-bell-table}
\end{table}

\section{Concluding Remarks}
\label{section4}
We designed and experimentally implemented a generalized
quantum scattering circuit to measure an $n$-point
correlation function on an NMR quantum information
processor, with an observable being measured sequentially at
these $n$ time instants.  We experimentally demonstrated the
violation of  a temporal noncontextuality PM inequality
using three NMR qubits, which involved performing sequential
noninvasive measurements.  We also demonstrated  the
violation of a transformed Bell-type inequality (analogous
to the temporal KCBS inequality) on the same system 
and showed that the Tsirelson bound of the transformed
Bell-type inequality is the same as that of the analogous
temporal KCBS inequality.  
The generalized quantum scattering circuit we have constructed is
independent of the quantum hardware used for its implementation and
can be applied to systems other than NMR qubits.
Our work asserts that NMR quantum
processors can serve as optimal test beds for testing such
inequalities.
\begin{acknowledgments}
All the experiments were performed on a Bruker Avance-III 600 MHz FT-NMR
spectrometer at the NMR Research Facility of IISER Mohali.  Arvind
acknowledges financial support from DST/ICPS/QuST/Theme-1/2019/General
Project number Q-68.  K.D. acknowledges financial support from
DST/ICPS/QuST/Theme-2/2019/General Project number Q-74.
\end{acknowledgments}
%
\end{document}